\documentclass[twocolumn,samsmath,amssymb,floatfix,superscriptaddress,nofootinbib]{revtex4}
\usepackage{graphicx}
\usepackage{dcolumn}
\begin{document}
\title{
Disentangling density and temperature effects \\ 
in the viscous slowing down of glassforming liquids}
\author{G.~Tarjus} 
\affiliation
{Laboratoire de Physique Th\'eorique des Liquides,
Universit\'e Pierre et Marie Curie, 4 place Jussieu, 
Paris 75005, France}
\author{D.~Kivelson}
\affiliation
{Department of Chemistry and Biochemistry, 
University of California, Los Angeles, California 90095}
\author{S.~Mossa}
\affiliation
{Laboratoire de Physique Th\'eorique des Liquides,
Universit\'e Pierre et Marie Curie, 4 place Jussieu, 
Paris 75005, France}
\author{C.~Alba-Simionesco}
\affiliation
{Laboratoire de Chimie Physique, B\^atiment 349, 
Universit\'e de Paris Sud, F-91405 Orsay, France}
\date{\today}
\begin{abstract}
We  present a consistent  picture  of the  respective role  of density
($\rho$)   and   temperature ($T$)   in   the  viscous  slowing  down  of
glassforming liquids and polymers. Specifically,  based in part upon a
new  analysis of    simulation  and  experimental   data   on   liquid
ortho-terphenyl, we conclude  that  a zeroth-order description of  the
approach to the  glass transition should be  formulated in terms of  a
temperature-driven  super-Arrhenius activated  behavior rather than  a
density-driven congestion or  jamming phenomenon. The  density plays a
role at a quantitative level, but its effect  on the viscosity and the
$\alpha$-relaxation time can be simply described via a single parameter, 
an effective interaction  energy that  is characteristic of  the high-$T$
liquid regime; as a result,  $\rho$ does not  affect the ``fragility'' of
the glassforming system.
\end{abstract}
\maketitle
\section{Introduction}
\label{intro:section}
Why should one care about the respective role  of the various external
control parameters, temperature   $T$, pressure $P$,   density $\rho$ (or
volume $V$) in  the viscous slowing down  of glass forming liquids and
polymers?  If the glass transition  at  $T_g$ were  a {\em bona  fide}
critical point with a diverging  correlation  length, the question  of
how  it is approached in   the $P-T$ or $\rho-T$   phase diagram would be
interesting but  not crucial  to determine the  main  features  of the
critical slowing down. However, there is no directly observed critical
point  associated   with    the  glass  transition  of   liquids   and
polymers.  For the canonical  fragile glassformer ortho-terphenyl, the
often  invoked  ``ideal  glass transition  temperature''   obtained by
extrapolating  the Vogel-Fulcher-Tammann expression  for the viscosity
and the $\alpha$ (or primary) relaxation time to infinity or extrapolating
the    excess    entropy   to    zero,   is   some   $40$    K   below
$T_g$~\cite{ediger96,tarjus01}.   Various  theories  predict ``avoided
critical points''    at   temperatures  above  $T_g$,  such   as   the
mode-coupling  singularity     at   $T_c$~\cite{goetze91}  or      the
frustration-limited-domain          avoided    transition           at
$T^*$~\cite{kivelson95},     or  ``unreachable critical   points''  at
temperatures below $T_g$, such    as  the ideal glass   transition  at
$T_o$~\cite{gibbs58,kirkpatrick89,mezard98},     but     whatever  the
approach, the  putative correlation  lengths  are expected  to remain of
modest        size,  no   more    than     $5$     to  $10$  molecular
diameters~\cite{ediger96,tarjus01}.  The  question of  which   control
parameter dominates the  slowing down  of  the relaxation may  then be
relevant to better understand the underlying physics.

Actually,   most  zeroth-order theoretical  descriptions  of  the slow
dynamics  of glass forming liquids   are formulated either at constant
temperature or at constant density. Two such limiting approaches are:

\noindent {\em 1)} the congestion, or jamming, picture of the glass
transition, in which slowing down  results  from the drainage of  free
volume  as   density increases~\cite{cohen64,liu98}.  The paradigm for
this approach is the hard-sphere fluid;

\noindent {\em 2)} a description in terms of thermally activated processes 
on a constant-density energy landscape~\cite{goldstein69,stillinger95}.   
In  this   approach   the landscape does  not change  as  one cools  
the system~\footnote{For  a given number  of particles $N$  and a given 
volume  $V$, hence a given density  $\rho=N/V$,   
the potential energy landscape, i.e., the
hypersurface of  potential energy plotted   as a function  of the $3N$
configurational coordinates, is independent  of temperature.}, and the
super-Arrhenius   $T$-dependence of  the  $\alpha$-relaxation  time and 
the viscosity is attributed to  changes    in the  minima and barriers
encountered  in  the exploration  of  the landscape  as one lowers the
temperature.

Of  course, such zeroth-order descriptions  can  always be improved to
account for some  additional dependence  on the ``secondary''  control
variable: temperature  can be included  in the  hard-sphere picture by
allowing the  diameter of the spheres to  change with temperature, and
the  energy landscape can be studied  for various  densities.  But the
mechanism   responsible for  the viscous   slowing  down is determined
predominantly by the density in the former case and the temperature in
the latter one.

In this article, we extend our previous
work~\cite{ferrer98,alba-simionesco02} to develop a consistent picture
of the roles of density and temperature.  To do so, we consider recent
studies  made by others as   well as new  simulation and  experimental
data.   The main   conclusion  is that  the   appropriate zeroth-order
description of  the slowing down of  relaxation  as one approaches the
glass transition is not that  of density (or volume)-driven congestion
or jamming, but that of a temperature-driven super-Arrhenius activated
behavior.   Density plays a  role at a quantitative  level: there is a
significant density dependence   of  the $\alpha$-relaxation time and the
viscosity at fixed temperature.  However, the density dependence can
be described in terms of a single interaction energy that is characteristic of
the high-T Arrhenius liquid. As a result,
the  temperature    dependence  of the  $\alpha$-relaxation time and the
viscosity  at    different   densities   collapse  onto    a   single,
material-dependent curve;  most importantly, this implies that density
does not affect the ``fragility'' of  a glass former, i.e., the degree
to which it exhibits super-Arrhenius temperature dependence.
\section{Assessing the influence of density and temperature}
\label{assessment:section}
Whereas physical pictures and  theories are conveniently formulated in
terms of temperature and density,  experiments are usually carried out
by  using  temperature and pressure   as control  variables.  Standard
studies   of the slowing   down of  relaxation   leading to the  glass
transition   are performed along    isobars, typically  at atmospheric
pressure,    by cooling the  liquid  or   the  polymer. The  resulting
$T$-dependence of, say, the $\alpha$-relaxation time $\tau_\alpha$ 
includes both an
intrinsic temperature effect  and a density  effect resulting from the
fact that cooling a system at constant  pressure usually increases its
density.    The temperature typically   decreases  by a factor  of $2$
between the normal  liquid  and the  glass transition,  whereas in the
same range  the density  increases typically   by only  $10\%$; but  of
course  those numbers are not meaningful  by themselves, the important
physical  information  being the  sensitivity of  the slowing  down to
changes of $T$ and $\rho$.

In order to disentangle  the  effects of  temperature and  density  at
constant pressure and provide a quantitative, model-free assessment of
their relative importance, we suggested some years ago~\cite{ferrer98}
to use  as a criterion the ratio  of  two coefficients of expansivity,
$|\alpha_\tau|/\alpha_P$, where  $\alpha_P=-\rho^{-1}(\partial \rho/\partial T)_P$  
is  the  usual isobaric expansivity and 
$\alpha_\tau=-\rho^{-1}(\partial \rho/\partial  T)_\tau$ is the  (unusual) isochronic
expansivity for a constant $\alpha$-relaxation time or a constant viscosity
($\eta$). Indeed,  along an isobar the  variation  with $T$ of $\ln(\tau_\alpha)$
where $\tau_\alpha$ is   expressed, say,  in   seconds (or $\ln(\eta)$  with  $\eta$
expressed, say, in Poise) can be decomposed as
\begin{equation}
\label{eq1:equation}
\frac{\partial \ln ( \tau_\alpha )}{\partial T}\bigg|_{P}=
\frac{\partial \ln ( \tau_\alpha )}{\partial T}\bigg|_{\rho}+
\frac{\partial \ln ( \tau_\alpha )}{\partial \rho}\bigg|_{T}
\frac{\partial \rho}{\partial T}\bigg|_{P},
\end{equation}
where the first term of the right-hand side is the intrinsic effect of
temperature at constant density  and the second one  is the  effect of
density    at  constant  temperature.   By  using  the above mentioned
coefficients of  expansivity, it is easy to  show that the former term
is  expressed in terms of  $\alpha_\tau$ and the latter  in terms of $\alpha_P$, so
that
\begin{equation}
\label{eq2:equation}
\frac{\partial \ln ( \tau_\alpha )}{\partial T}\bigg|_{P}=
\frac{\partial \ln ( \tau_\alpha )}{\partial \ln ( \rho )}\bigg|_{T}
(-\alpha_\tau+\alpha_P),
\end{equation}
and   the  ratio  $|\alpha_\tau|/\alpha_P$  measures  the   relative importance  of
temperature  versus   density contributions at  a  given thermodynamic
point (recall that in general $\alpha_P>0$  and $\alpha_\tau<0$): 
if $|\alpha_\tau|/\alpha_P>1$,
temperature is a  more   important controlling factor of   the viscous
slowing down than density.

Note that an equivalent  characterization of the respective weights of
temperature and    density,  although  not    expressed  in  terms  of
thermodynamic coefficients of  expansivity, was proposed by Naoki {\em
et al.}~\cite{naoki87}.    These authors suggested  use  of  the ratio
$E_V/H_P$, where          $E_V=(\partial\ln(\tau_\alpha)/\partial(1/T))_\rho$       
and
$H_P=(\partial\ln(\tau_\alpha)/\partial(1/T))_P$. It is easy to show that
\begin{equation}
\label{eq3:equation}
\frac{E_V}{H_P}=\frac{|\alpha_\tau|}{|\alpha_\tau|+\alpha_P},
\end{equation}
so that, when $|\alpha_\tau|/\alpha_P>1$, $E_V/H_P>1/2$.
\begin{figure}[t]
\centering
\includegraphics[width=0.45\textwidth]{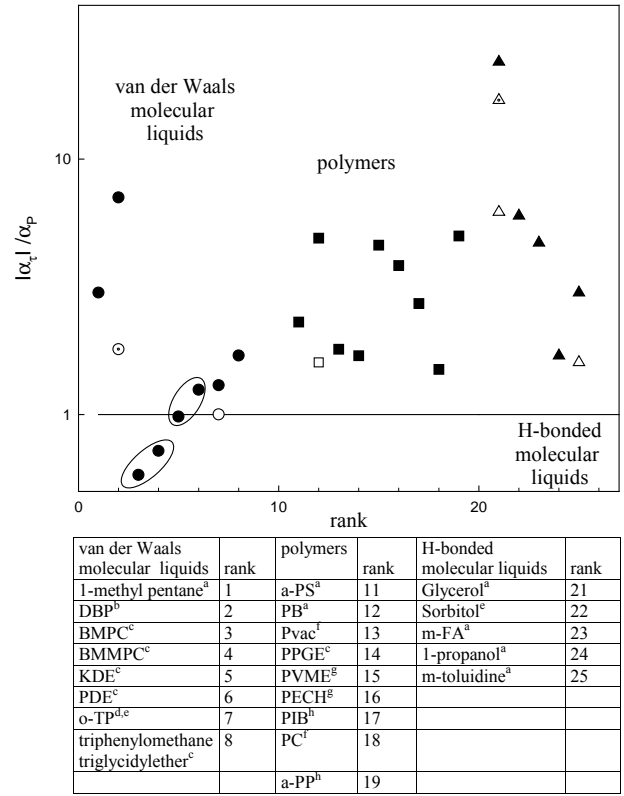}
\caption{Ratio $|\alpha_\tau|/\alpha_P$ for various  liquids  and 
polymers evaluated  at the glass transition and at higher  
temperature (at atmospheric pressure): filled symbols  for  $T_g$ 
(see  text), dotted open  symbols for $\tau_\alpha\sim
10^{-5}$s or  $\tau_\alpha\sim 10^{-1}$ s , open  symbols for 
$\tau_\alpha\sim 10^{-9}$ s or $\eta\sim 10^4$  cPoise.  
The two  ellipses indicate points corresponding to
liquids of the same family with a very similar chemical structure. The
point for  DBP  at Tg  should be  taken  with  caution because of  the
inaccuracy of the equation of state.  The rank of the liquids (see the
table) within the three   classes,  van der Waals  molecular  liquids,
polymers,    and H-bonded     molecular    liquids,   is   arbitrarily
chosen.   References for  the   table:  (a)   Ref.~\cite{ferrer98} and
references   therein, (b) Refs.~\cite{fujimori97}, ~\cite{cook93}  and
~\cite{cook94}, (c) Ref.~\cite{paluch02}, (d) Ref.~\cite{schug98}, (e)
Ref.~\cite{naoki87}, (f) Ref.~\cite{huang02},  (g) Ref.~\cite{aude03},
(h) Ref.~\cite{chauty03}, (i) Ref.~\cite{hollander01}.}
\label{fig1}
\end{figure}

From available experimental  data on viscosity and $\alpha$-relaxation time
we have calculated (or collected)  the ratio $|\alpha_\tau|/\alpha_P$ for a variety
of glassforming liquids and polymers interacting through van der Waals
forces or hydrogen   bonds.  Strong, network-forming systems are   not
considered  here~\footnote{  It is    likely  that for  such  strongly
interacting glassformers temperature is the dominant driving parameter
of the  viscous slowing down so long  as pressure does not  change the
local structure of  the   liquid.}.  We show in   Fig.~\ref{fig1}  the
ratios calculated at  atmospheric pressure  (i)  at $T_g$ (either  the
calorimetric $T_g$ corresponding to an isochrone with $\tau_\alpha\sim 1000$ s or
the dielectric $T_g$ with $\tau_\alpha\sim 1-100$ s) and (ii), when possible, for
a    smaller  $\alpha$-relaxation time   or    viscosity  (i.e., a   higher
temperature).   It  is quite  noticeable  that  the ratio at  $T_g$ is
significantly larger than  $1$, with   the exception  of a  family  of
organic liquids with very similar chemical structure (BMPC, BMMPC) for
which it is slightly less  than $1$~\cite{paluch02}. The largest value
is for the $H$-bonded liquid glycerol; van der Waals molecular liquids
tend  to have smaller ratios than  polymers and  H-bonded liquids, but
the correlation between  the value of the ratio  and the nature of the
glassforming system is not very strong.

A   few additional comments   are worth  making.  First, there  may be
significant  errors  in  determining $\alpha_\tau$ and $\alpha_P$: 
high-precision
thermodynamic and dynamic data  are needed to obtain such second-order
quantities.  Secondly, and  maybe somehow contrary to the  expectation
that density becomes more  important as pressure increases, the  ratio
$|\alpha_\tau|/\alpha_P$ is found to increase  with pressure 
(except for glycerol),
at least up to  a  few kbars. Finally,  the   ratio decreases as   the
temperature increases or, equivalently, as $\tau_\alpha$ (or $\eta$)
decreases. It   is indeed expected  that  in the  ``normal'', high-$T$
liquid density and temperature  play more  comparable roles, at  least
for weakly interacting systems.  However, this  is rather at odds with
the congestion/jamming picture of   the glass transition in  which one
would expect density to become predominant as one gets close to $T_g$.

To  summarize  the  results   of this  section:  overall,  temperature
dominates  over  density   in driving  the  viscous slowing   down  of
glassforming liquids and polymers;  for a quantitative description one
cannot bluntly neglect the effect of density;  the role of temperature
becomes more important in the super-Arrhenius regime as one approaches
the glass transition.
\section{Scaling out the density dependence}
\label{scaling:section}
The conclusions of   the  preceding section, together with    previous
work~\cite{ferrer98,alba-simionesco02},  lead us to  describe the  $\alpha$
relaxation   in supercooled  liquids  and   in polymers as   thermally
activated,  i.e., for practical purposes  describable by the following
$T$-dependence:
\begin{equation}
\label{eq4:equation}
\tau_\alpha(\rho,T)=\tau_\alpha^\infty(\rho)
\exp\left[E(\rho,T)/T\right],
\end{equation}
where  the effective activation  energy $E(\rho,T)$ is  at least equal to
several  times the  thermal energy  $T$  (we work  in  units such that
$k_B=1$).  A zeroth-order physical picture of the viscous slowing down
of fragile glass formers and   of  their crossover from Arrhenius   to
super-Arrhenius behavior should thus focus on the role of temperature,
the effect of density being  incorporated in quantities characterizing
the    high-$T$,   ``normal''  liquid   phase.   As   we  suggested in
Ref.~\cite{alba-simionesco02},  the  simplest     description  is that
density only  enters via a  characteristic interaction energy that one
can associate  with the energy barrier   to $\alpha$-relaxation and viscous
flow in the high-$T$ Arrhenius-like regime, $E_\infty(\rho)$; this energy sets
the scale for both  the effective activation  energy $E(\rho,T)$  and the
temperature.  Accordingly, all  isochoric data should collapse  onto a
master curve
\begin{equation}
\label{eq5:equation}
\frac{E(\rho,T)}{E_\infty(\rho)}=\Phi\left(\frac{T}{E_\infty(\rho)} \right),
\end{equation}
where $\Phi(\zeta)$  is a monotonously  decreasing  function of $\zeta$  that  is
equal to  $1$  for $\zeta$ larger than   some crossover value $\zeta^*$.   If,
moreover,   the   $\rho$-dependence   of    the  prefactor    $\tau_\alpha^\infty$  in
Eq.~(\ref{eq4:equation})   is  negligible, one   can  also  obtain   a
mastercurve for the $\alpha$-relaxation time itself,
\begin{equation}
\label{eq6:equation}
\ln\left(\tau_\alpha(\rho,T)\right)=\ln(\tau_\alpha^\infty)+
\frac{E_\infty(\rho)}{T}\Phi\left(\frac{T}{E_\infty(\rho)}\right),
\end{equation}
where $\tau_\alpha(\rho,T)$and $\tau_\alpha^\infty$ are both  expressed, 
say, in seconds. It is
clear from Eqs.~(\ref{eq5:equation}) and (\ref{eq6:equation}) that the
effect of density   could be {\em  quantitatively} significant without
{\em qualitatively} determining the slowing  down of the relaxation as
one approaches the glass transition.
\begin{figure}[t]
\centering
\includegraphics[angle=-90,width=0.45\textwidth]{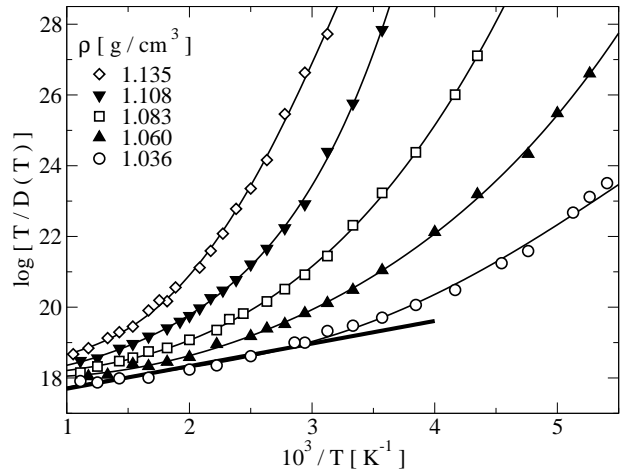}
\caption{Simulation data on a   model ortho-terphenyl: $\ln(T/D(\rho,T))$,   where
$D(\rho,T))$  is the translational diffusion   constant, versus $1/T$ for
several values of $\rho$ (from top to bottom, $\rho=1.135,1.108,1.083,1.060,
1.036$  $g/cm^3$).  The  straight  line is  an  Arrhenius  fit to  the
high-temperature portion  of the data ($T\leq T^*(\rho)$),   shown   here for
$\rho=1.036$ $g/cm^3$.}
\label{fig2}
\end{figure}

In Ref.~\cite{alba-simionesco02} we gave two examples of data collapse
onto the mastercurve  in Eq.~(\ref{eq5:equation})  (for the simulation
data on the   binary  Lennard-Jones mixture~\cite{sastry00} and    for
experimental  data  on glycerol~\cite{johari73,cook94}).  It  is worth
stressing  that, despite  decades of pressure  studies of glassforming
liquids and  polymers, the available  data  base of isochoric data  is
scant. In  what follows we  present new  simulation   data on a  model
ortho-terphenyl and, building upon  the work  of Refs.~\cite{toelle01}
and~\cite{dreyfus03}  we     also   analyze   experimental    data  of
ortho-terphenyl.
\begin{figure}[t]
\centering
\includegraphics[width=0.45\textwidth]{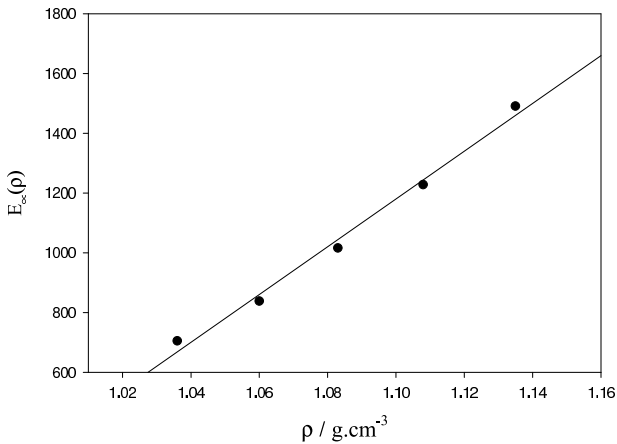}
\caption{High-$T$ Arrhenius activation energy $E_\infty$ versus 
density for the data shown
in Fig.~\protect\ref{fig2}.}
\label{fig3}
\end{figure}
\begin{figure}[t]
\centering
\includegraphics[width=0.5\textwidth]{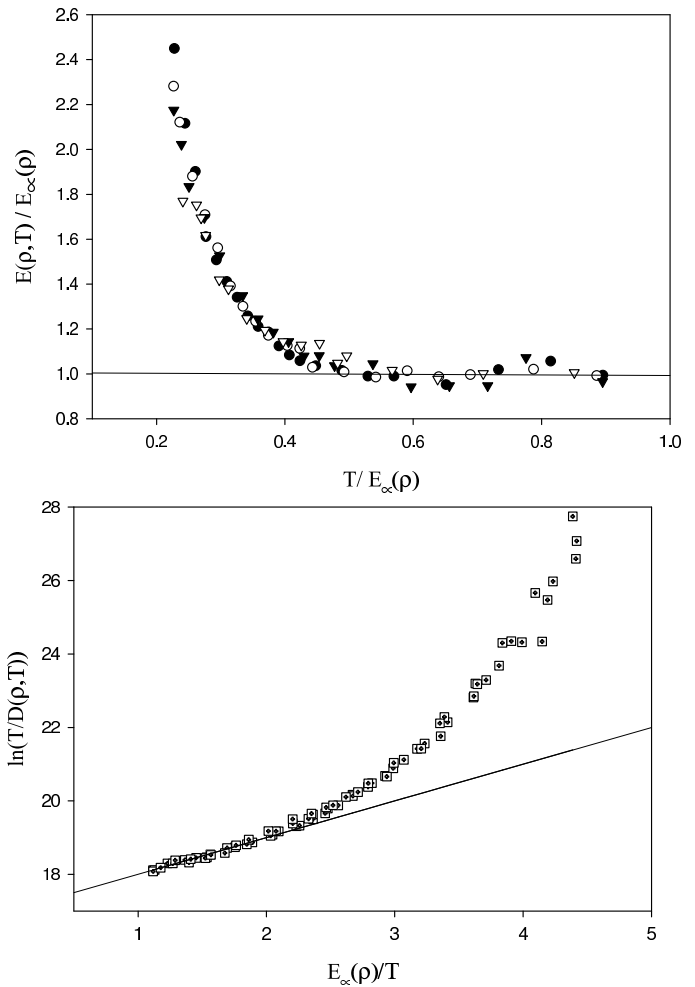}
\caption{Mastercurves for the simulation data on model ortho-terphenyl
at $5$ different densities:
{\em Top:} $E(\rho,T)/E_\infty(\rho)$ versus $T/E_\infty(\rho)$;
{\em Bottom:} $ln(T/D(\rho,T))$ versus $E_\infty(\rho)/T$. 
Symbols as in  Fig.~\protect\ref{fig2}.}
\label{fig4}
\end{figure}
\section{Analysis of simulation and experimental data on ortho-terphenyl}
\label{data:section}
We have considered the simulation data  of Ref.~\cite{mossa02} for the
now        widely        studied       Lewis-Wahnstr\"om    model     of
ortho-terphenyl~\cite{wahnstroem93},   a  model   of  rigid three-site
molecules  with intermolecular  site-site Lennard-Jones  interactions.
In view of the analysis below, the data base  of state points has been
further extended in the  high-temperature region.  Simulation  details
are  given in    Ref.~\cite{mossa02}.   The  translational   diffusion
constant $D(\rho,T)$  has been considered for  $5$ different densities in
the range $1.036$ to  $1.135$ g/cm$^3$  (representing a $10\%$  overall
change in density)
\footnote{Note that, in the range of temperature accessible to our
computer simulations, there is  no significant decoupling between  the
translational diffusion constant and the other relaxation times.}.  To
obtain data  more  akin to  viscosity   or  relaxation time,  we  have
considered  $T/D(\rho,T)$: see Fig.~\ref{fig2}.   From the Arrhenius plot
of  Fig.~\ref{fig2},  we  have determined  for   each curve a high-$T$
activation  energy $E_\infty(\rho)$   that  describes  the  data   above  some
crossover    temperature $T^*(\rho)$,  below   which the  curves markedly
deviate  from  an Arrhenius  dependence.   The values  of $E_\infty(\rho)$ are
shown in Fig.~\ref{fig3}.  As can   be seen from Fig.~\ref{fig4},  the
data  taken along the $5$  different  isochores collapse onto a single
curve when  energies  and temperatures  are rescaled  by the parameter
$E_\infty(\rho)$; the prefactor $\tau_\alpha^\infty$ being independent 
of density, a master
curve   is obtained      by using  either     Eq.~(\ref{eq5:equation})
(Fig.~\ref{fig4} Top) or Eq.~(\ref{eq6:equation}) (Fig.~\ref{fig4} Bottom).

\begin{figure}[t]
\centering
\includegraphics[width=0.45\textwidth]{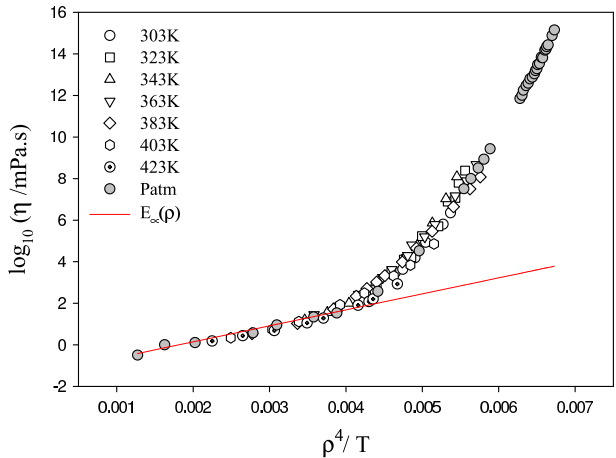}
\caption{Mastercurve experimental data on the viscosity of liquid 
ortho-terphenyl: $\log_{10}(\eta)$ versus $\rho^4/T$. 
The data at atmospheric pressure are taken 
from Refs.~\protect~\cite{laughlin72,greet67} and those at higher 
pressures from Ref.~\protect~\cite{schug98} 
(only pressures up to $5$ kbars are considered). 
The straight line is the best Arrhenius fit to the 
high-$T$/low-$\rho$ portion of the data.}
\label{fig5}
\end{figure}

Computer simulation offers a convenient means  to study isochoric data
directly, but     the range of   temperature  and  of  relaxation time
attainable   is quite limited.    Ideally, isochoric experimental data
would provide a  more stringent test  for Eqs.~(\ref{eq5:equation}) or
(\ref{eq6:equation}).   However,  even   for the  well-studied  liquid
ortho-terphenyl, obtaining isochoric data   for the viscosity  or  the
$\alpha$-relaxation time requires significant  manipulation of the original
data (collected along isobars and isotherms).  We follow here the work
of others~\cite{dreyfus03} who  gave evidence that  the $\alpha$-relaxation
time of ortho-terphenyl for a large range of pressure and temperature,
hence of density and temperature, can be  placed to a good accuracy on
a  single master  curve when  it  is plotted  as   a function  of $\rho\,
T^{-1/4}$:  see, e.g.,   Fig.~11  of Ref.~\cite{dreyfus03}.   We  have
repeated a similar analysis for the  viscosity, extending the range to
higher   temperatures by  using   the      data  of Schug  {\em     et
al.}~\cite{schug98}. The resulting mastercurve, $\ln(\eta)$ versus $\rho^4 /
T$ is shown in Fig.~\ref{fig5} (the inset shows a similar plot for the
$\alpha$-relaxation time).  Data collapse is very  good. One should however
keep in  mind a few words  of caution.  First,  there is a significant
discrepancy  between   $\alpha$-relaxation  times obtained    by  different
techniques.   Secondly, we have included   data only up  to $5$  kbars
(similarly   in Ref.~\cite{dreyfus03} only data up   to $2$ kbars were
considered); for higher pressures, the viscosity data start to deviate
from the  mastercurve,  which may  be due  to  a much  more  uncertain
equation of state, to experimental errors, and/or to the fact that the
$\rho\, T^{-1/4}$ scaling somewhat breaks down at very high pressures.

What   is   the   meaning  of  the   $\rho\, T^{-1/4}$ scaling? In
Ref.~\cite{toelle01} it was suggested that this scaling is reminiscent
of  that occurring  in the  fluid of  soft  spheres interacting via an
$r^{-12}$ potential. Indeed for particles  interacting via a power-law
$r^{-n}$ pair potential,  the $\rho$  and  $T$ dependences of  all excess
thermodynamic properties \footnote{An excess thermodynamic quantity is
defined here as the contribution due to the interactions that comes on
top of an  ideal-gas result.}  enter only  through a single  parameter
$\Gamma=\rho\,  T^{-3/n}$\cite{hansen86}; the relaxation  time is then 
essentially a function
of   $\Gamma$  whereas    the   equation of    state   takes  the   form $P
T^{-(1+3/n)}={\cal F}(\rho\,  T^{-3/n})$~.  It  would  be,
however, somewhat   surprising    if  a  molecular  liquid  such    as
ortho-terphenyl  behaved like a fluid of  soft repulsive spheres (with
$n=12$  as in a Lennard-Jones  potential).  Actually, from the data in
the literature it  is easy to check that   the equation of state  does
{\em not}   obey    the predicted  form   $P\, T^{-5/4}={\cal   F}(\rho\,
T^{-1/4})$.
\begin{figure}[t]
\centering
\includegraphics[width=0.45\textwidth]{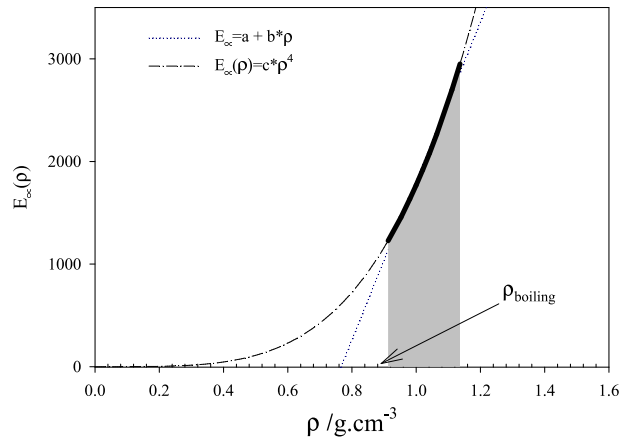}
\caption{High-$T$  Arrhenius activation energy  $E_\infty$  versus density from  the
viscosity  data  of  ortho-terphenyl. The    dot-dashed curve  is  the
$\rho^4$-dependence derived from Fig.~\protect\ref{fig5}; the dotted line
is a  linear fit to the $\rho^4$-curve  over the relevant range of liquid
densities (shaded area).}
\label{fig6}
\end{figure}

We  suggest instead that the empirical  scaling found for the viscosity
and    the  $\alpha$-relaxation  time  of  ortho-terphenyl    is  one  more
manifestation  of the mastercurve discussed  in the preceding section:
the  data collapse  shown  in  Fig.~\ref{fig5}  is  a  special case of
Eq.~(\ref{eq6:equation}) with $E_\infty(\rho)\propto\rho^4$.  One  can indeed  see from
Fig.~\ref{fig5}  that the  small-$(1/\zeta)$  portion of  the curve, where
$1/\zeta=\rho^4/T$, is  linear to a  good approximation, which corresponds to
an Arrhenius  behavior, $\ln(\eta(\rho,T))=\ln(\eta_\infty)+E_\infty(\rho)/T$, 
with $E_\infty(\rho)$
proportional  to $\rho^4$.  In  Fig.~\ref{fig6} we  have plotted $E_\infty(\rho)$
for   the range of densities   actually  studied experimentally: it is
obvious from the figure that for this range of densities, a linear fit
to $E_\infty(\rho)$ would do just as well as the power law $\rho^4$.  This latter
may then  just   be a  coincidence  without  any  substantial physical
meaning.   On the other hand, the  observed data collapse supports our
claim  that  the influence  of the   density on  the slow  dynamics of
glassformers  can be reduced to  a single effective interaction energy
$E_\infty(\rho)$.
\section{Comparison between systems: relative ``fragilities''}
\label{fragilities:section}
\begin{figure}[t]
\centering
\includegraphics[width=0.45\textwidth]{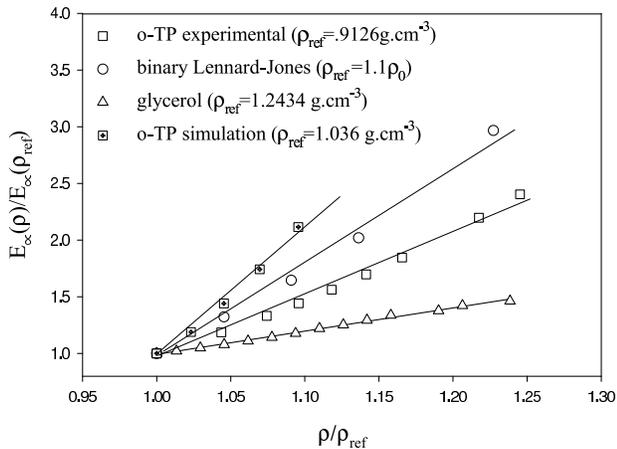}
\caption{$E_\infty(\rho)/E_\infty(\rho_{ref})$,  where  $\rho_{ref}$  
is a small  reference liquid
density, for the glassformers: binary Lennard-Jones mixture, glycerol,
Lewis-Wahnstr\"om          model       of     ortho-terphenyl       (see
Fig.~\protect\ref{fig3}),      and        ortho-terphenyl         (see
Fig.~\protect\ref{fig6}).}
\label{fig7}
\end{figure}

\begin{figure}[t]
\centering
\includegraphics[width=0.5\textwidth]{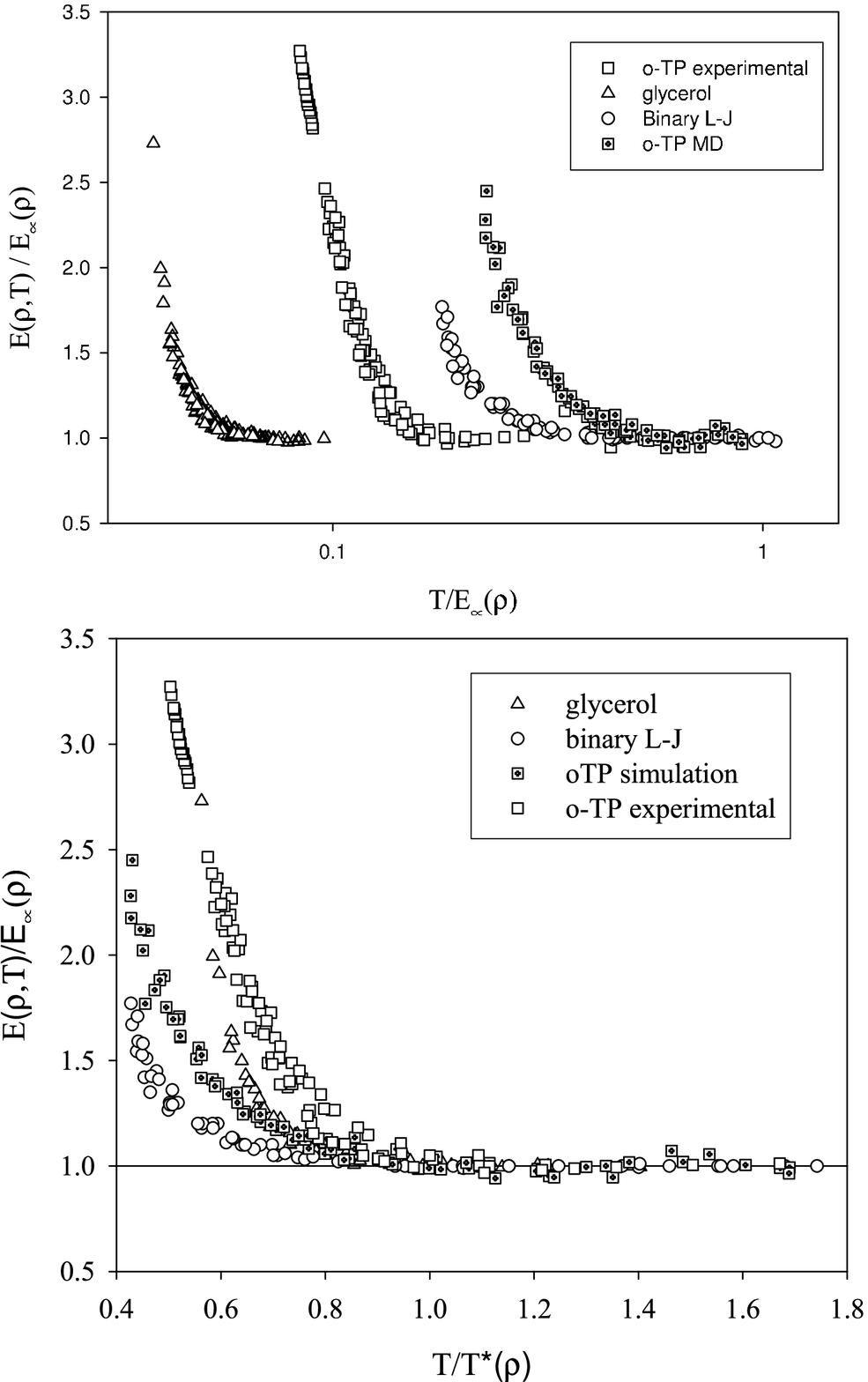}
\caption{$E(\rho,T)/E_\infty(\rho)$ versus (Top) $\ln(T/E_\infty(\rho))$ and 
(Bottom) $T/T^*(\rho)$ for the
$4$ glassformers indicated in the text and in Fig.~\protect\ref{fig7};
$T^*(\rho)=\zeta_*  E_\infty(\rho)$ is the   crossover  temperature above which   the
$T$-dependence of $\eta$ or $\tau_\alpha$ is essentially Arrhenius-like. $\zeta_*$ is
species-specific:  $\zeta_*=0.42$ for  the binary  Lennard-Jones  mixture;
$\zeta_*=0.075$  for glycerol; $\zeta_*=0.53$  for  the model ortho-terphenyl;
$\zeta_*=17$ for real ortho-terphenyl.}
\label{fig8}
\end{figure}

The scaling function $\Phi(\zeta)$  appearing  in the mastercurves given   in
Eqs.~(\ref{eq5:equation})       and   (\ref{eq6:equation})          is
species-specific, as are  the $\rho$-dependence and  the magnitude of the
effective  interaction  energy   $E_\infty(\rho)$.   This  is  illustrated  in
Figs.~\ref{fig7}  and \ref{fig8} for  the  two systems studied in  the
present work, computer simulated and real  ortho-terphenyl, as well as
for    the two systems   considered in  a   previous study, the binary
Lennard-Jones mixture and (real) glycerol.  In Fig.~\ref{fig7} we have
plotted $E_\infty(\rho)/E_\infty(\rho_{ref})$ versus $\rho/\rho_{ref}$, 
where $\rho_{ref}$ is a
liquid density taken as a reference (e.g., the  density at the boiling
point when it is known), for the four glassformers.  One observes that
besides the fact that over the range considered  (a variation of about
$10\%$)  the $\rho$-dependence is essentially linear,  the four curves are
all      different~\footnote{This,  as   well   as    the   results in
Fig.~\ref{fig8},  casts  some    doubts    on the   ability   of   the
Lewis-Wahnstr\"om     model     to    accurately    represent     liquid
ortho-terphenyl.}.

In Fig.~\ref{fig8}    we  show   the  curves   $E(\rho,T)/E_\infty(\rho)$  versus
$\ln(T/E_\infty(\rho))$ and  $T/T^*(\rho)$, where    $T^*(\rho)=\zeta_* E_\infty(\rho)$ 
is   the
crossover   temperature     below  which  a    marked  super-Arrhenius
$T$-dependence   is observed;   in    terms   of    mastercurves  (see
Eq.~(\ref{eq5:equation})), this    amounts to plotting   $\Phi(\zeta)$ versus
$\zeta/\zeta_*$ where  $\Phi(\zeta)\simeq1$ for $\zeta\leq \zeta_*$,  $\zeta_*$ 
being a species-dependent
constant.  The  differences observed  for  $\zeta > \zeta_*$ among  the four
systems  reflect the different   degrees of super-Arrhenius  behavior,
i.e., the different  fragilities: one finds,  as expected, that (real)
ortho-terphenyl  and glycerol are  significantly more fragile than the
two  computer  simulated systems  (the  highest  points in  activation
energy       for      glycerol    should     be          taken    with
caution~\cite{alba-simionesco02}).

It is worth pointing out that a consequence of the existence of the 
mastercurves is that {\em fragility is independent of density}.
This applies no matter what particular definition of (dynamic) 
fragility~\cite{angell91} one chooses. Consider, for instance, the 
commonly used steepness index, 
\begin{equation}
\label{eq8:equation}
m_\rho=\frac{\partial\ln(\tau_\alpha(\rho,T))}
{\partial(T_g(\rho)/T)}\bigg|_{T_g(\rho)},
\end{equation}
evaluated  along an  isochoric path. Applying Eq.~(\ref{eq6:equation})
and defining as usual $T_g(\rho)$ as the temperature at which $\tau_\alpha(\rho,T)$,
expressed in seconds, is  equal to a given value, $\ln(\tau_\alpha(\rho,T))={\cal
K}$, one finds that
\begin{equation}
\label{eq9:equation}
m_\rho=\frac{E_\infty(\rho)}{T_g(\rho)}\frac{\partial\ln(\tau_\alpha(\rho,T))}
{\partial(T_g(\rho)/T)}\bigg|_{T_g(\rho)}=-\zeta_g \frac{d(\Phi(\zeta)/\zeta)}
{d\zeta}\bigg|_{\zeta_g},
\end{equation}
where $\zeta_g=T_g(\rho)/E_\infty(\rho)$ and $d(\Phi(\zeta)/\zeta)/d\zeta$ 
is negative. According to
the  above  definition  of $T_g(\rho)$   and to Eq.~(\ref{eq6:equation}),
$\zeta_g$  is simply  the   solution of the  equation  $(\Phi(\zeta_g)/\zeta_g)={\cal
K-\ln(\tau_\alpha^\infty)}$, and is thus independent of $\rho$.  As a result, whatever
the value of ${\cal K}$ chosen, the steepness index $m$ is independent
of  the density.   This remains  true if the  prefactor $\tau_\alpha^\infty$  has a
small  residual $\rho$-dependence.    The opposite conclusion  reached in
Ref.~\cite{sastry01}  from data  on  the binary  Lennard-Jones mixture
probably  arises from the large  intrinsic  inaccuracy associated with
fitting a $3$-parameter Vogel-Fulcher-Tammann formula to a non-fragile
system (see Fig.~\ref{fig8}) over a limited time-scale domain when the
putative  $T_o$  is   at  much  lower temperature   than accessible by
computer simulation.

The fact that   fragility is independent  of  density does  not imply,
however, that it is independent of  pressure. Fragility measured along
isobaric paths  involves  additional   effects, associated with    the
variation with   $\rho$ of $E_\infty$   and  of the  isobaric  coefficient  of
expansivity $\alpha_P$.  As a result,  fragility may increase, decrease, or
be  non-monotonic as a function   of pressure.  We will discuss  this
point in more detail in a forthcoming publication.
\section{Conclusion}
\label{conclusion:section}
Extending  our earlier work we  have clarified  the respective role of
density  and temperature in the viscous  slowing  down of glassforming
liquids and polymers.  On the basis of available experimental data, we
have argued  that  this slowing    down  is best described   {\em   at
zeroth-order}    in terms of     thermally  activated processes  whose
characteristic super-Arrhenius behavior is  not primarily driven by  a
congestion or   jamming  phenomenon resulting,  for  instance,  from a
drainage of  free volume as one  approaches the  glass transition, but
should   rather be   explained  as  an   intrinsic temperature  effect
operating at constant  density.   Although density effects   cannot be
neglected  {\em at a  quantitative level}, we  have brought additional
evidence showing that   these effects can  be  described via  a single
effective  interaction   energy   $E_\infty(\rho)$  that  characterizes    
the relaxation of the liquid in its high-temperature range. In the range
of  accessible liquid densities,  the variation of   $E_\infty$ with $\rho$ is
essentially linear and featureless.

As a result of  the collapse of  the viscosity and $\alpha$-relaxation time
data obtained  at various densities onto  a mastercurve, we have shown
that  the (dynamic) fragility  of   glassforming liquids, i.e.,  their
degree of     departure from  the   high-$T$ Arrhenius   behavior,  is
independent of density.  This, in  turn, supports the above conclusion
that  a  physical explanation of the   super-Arrhenius behavior of the
viscosity  and   $\alpha$-relaxation   time as one    approaches the  glass
transition   should be found in   the intrinsic effect of temperature.
Although we feel that we have  provided strong evidence to support our
conclusions, is should again be stressed that there is a need for more
extensive sets of experimental data and accurate equations of state to
test on   a wider scale  the proposed   scaling form  of  the  density
dependence of the viscosity and the $\alpha$-relaxation time.
\begin{acknowledgments}
LPTL and LPC are UMR 7600 and 8000, respectively, at the CNRS.
\end{acknowledgments}
\end{document}